\documentclass[epj]{webofc}\usepackage[varg]{txfonts}\usepackage{bm}
\woctitle{QCD@Work 2014}
\begin{document}\title{Semirelativistic Bound-State Equations:
Trivial Considerations}\author{Wolfgang Lucha\inst{1}\fnsep
\thanks{\email{Wolfgang.Lucha@oeaw.ac.at}}\and Franz F.~Sch\"oberl
\inst{2}\fnsep\thanks{\email{franz.schoeberl@univie.ac.at}}}
\institute{Institute for High Energy Physics, Austrian Academy of
Sciences, Nikolsdorfergasse 18, A-1050 Vienna, Austria\and Faculty
of Physics, University of Vienna, Boltzmanngasse 5, A-1090 Vienna,
Austria}\abstract{Observing renewed interest in long-standing
(semi-) relativistic descriptions~of two-body bound states, we
would like to make a few comments on the eigenvalue problem posed
by the spinless Salpeter equation and, illustrated by the examples
of the nonsingular Woods--Saxon potential and the singular
Hulth\'en potential, recall elementary tools that, in their quest,
practitioners looking for analytic albeit approximate solutions
will find~useful.}\maketitle

\section{Introduction: Spinless Salpeter equation}\label{Sec:ISEE}
Recent years have witnessed a rise of attempts to study bound
states by (semi-) relativistic equations of motion, such as the
Klein--Gordon equation, the Dirac equation, or (as a
straightforward generalization of the Schr\"odinger equation) the
\emph{spinless Salpeter equation\/}, with all its merits and
drawbacks (consult, for instance,
Refs.~\cite{Lucha:SSEa,Lucha:SSEb} for details), derived by
nonrelativistic reduction (cf., for instance,
Refs.~\cite{Lucha:NRa,Lucha:NRb,Lucha:NRc}) of the Bethe--Salpeter
equation \cite{BSE,SE}. For two particles of (just for notational
simplicity) equal masses, $m,$ and relative momentum $\bm{p},$
interacting via a potential $V(\bm{x})$ depending on their
relative coordinate, $\bm{x},$ the spinless Salpeter equation may
be regarded as the eigenvalue equation of the nonlocal Hamiltonian
\begin{equation}H\equiv T(\bm{p})+V(\bm{x})\ ,\qquad
T(\bm{p})\equiv2\sqrt{\bm{p}^2+m^2}\ ,\label{Eq:H}\end{equation}
incorporating the relativistic kinetic energy, $T(\bm{p}).$ In
view of the interest noted, we revisit this equation for central
potentials $V(\bm{x})=V(r),$ $r\equiv|\bm{x}|,$ by recalling (and
exploiting) a couple of well-known results. More precisely, in an
almost telegraphic style we sketch, in Sect.~\ref{Sec:ASSC}, some
issues relevant for relativistic quantum theory and apply the
insights gained, in Sects.~\ref{Sec:AWSP} and \ref{Sec:AHP}, to
nonsingular and singular potentials.

\section{Approximate solutions: Strict constraints}\label{Sec:ASSC}
\subsection{Existential question: Maximum number of bound states
that can be accommodated}\label{Subsec:TEQ}In contrast to the
Coulomb potential $V_{\rm C}(r)\equiv-\kappa/r,$ $\kappa>0,$ lots
of rather popular potentials (for instance, the Yukawa or the
Woods--Saxon potential) admit only a \emph{finite\/} number, $N,$
of bound states: this number is a crucial characteristic of
bound-state problems. For generic (nonrelativistic)
\emph{Schr\"odinger\/} operators\begin{equation}H_{\rm
NR}\equiv\frac{\bm{p}^2}{2\,\mu}+V(r)\ ,\qquad\mu>0\ ,\qquad
V(r)\le0\ ,\label{Eq:Schreq}\end{equation}with reduced mass $\mu$,
the perhaps most easy-to-evaluate upper bound to $N$ is that one
by
Bargmann~\cite{Bargmann}:\begin{equation}N\lneqq\frac{I\,(I+1)}{2}\
,\qquad I\equiv2\,\mu\int\limits_0^\infty{\rm d}r\,r\,|V(r)|\
.\end{equation}For \emph{semirelativistic\/} Hamiltonians of the
spinless-Salpeter form (\ref{Eq:H}), an upper bound to $N$ is
given~by~\cite{Daubechies}
\begin{equation}N\le\frac{C}{12\,\pi}\int_0^\infty\limits{\rm
d}r\,r^2 \left[|V(r)|\left(|V(r)|+4\,m\right)\right]^{3/2}\
,\qquad\!\begin{array}{l}C=6.074898\qquad(m=0)\
,\\C=14.10759\qquad(m>0)\ .\end{array}\end{equation}

\subsection{Narrowing down solutions: Rigorous bounds on
eigenvalues}\label{Subsec:NDS}As a function of $\bm{p}^2,$
$T(\bm{p})$ is \emph{concave\/}. Thus, $H$ is bounded from above
by its \emph{Schr\"odinger\/} limit
\cite{Lucha:SchrUBa,Lucha:SchrUBb}:
\begin{equation}H\le2\,m+\frac{\bm{p}^2}{m}+V(\bm{x})\
.\end{equation}The Rayleigh--Ritz \emph{variational\/} technique
applies to self-adjoint (Hilbert-space) operators, $H,$ bounded
from below, with eigenvalues $E_0\le E_1\le E_2\le\cdots$:
\emph{The $d$ likewise ordered eigenvalues of $H,$ restricted to
any trial subspace of dimension $d$ of the domain of $H$, form
upper bounds to the lowest $d$ eigenvalues of $H$ below the onset
of its essential spectrum.} It is favourable to know one's
preferred basis of this trial space analytically in \emph{both\/}
position and momentum spaces. We can achieve this by choosing
\cite{LagBda,LagBdb} an orthonormal basis defined by means of
generalized-Laguerre polynomials $L_k^{(\gamma)}(x)$
\cite{Abramowitz} for parameter $\gamma,$ utilizing two
variational parameters, $\mu$ (with unit mass dimension) and
$\beta$ (which is~dimensionless), and spherical harmonics ${\cal
Y}_{\ell m}(\Omega)$ of angular momentum $\ell$ and projection $m$
depending on the solid~angle~$\Omega$:
\begin{align}&\psi_{k,\ell m}(\bm{x})\propto r^{\ell+\beta-1}
\exp(-\mu\,r)\, L_k^{(2\ell+2\beta)}(2\,\mu\,r)\,{\cal Y}_{\ell
m}(\Omega)\ ,\label{Eq:LB}\\&L_k^{(\gamma)}(x)\equiv
\sum_{t=0}^k\,(-1)^t\,\binom{k+\gamma}{k-t}\,\frac{x^t}{t!}\
,\qquad k=0,1,2,\dots\ .\end{align}For the \emph{lower\/} end of
the spectrum of $H,$ the operator inequality
$T(\bm{p})\ge2\,m\ge0$ implies $E_0\ge\inf_{\bm{x}}V(\bm{x}).$

\subsection{Boundedness from below: Constraints on potential
parameters}\label{Subsec:BfB}As an even positive operator, the
kinetic-energy term $T(\bm{p})$ is definitely bounded from below.
However, for a potential $V(\bm{x})$ that is not bounded from
below, the issue of the boundedness from below of~the full
Hamiltonian (\ref{Eq:H}) has to be addressed: The operator $H$
might turn out to be bounded from below~only for
crucial-potential-parameter values within adequate ranges. For the
semirelativistic Coulomb problem, this question has been nicely
answered by Herbst a long time ago \cite{Herbst}. In general, this
question~may be discussed by deriving upper bounds to energy
levels, in particular, to the ground state, by using the trial
states (\ref{Eq:LB}) for quantum numbers $k=\ell=m=0$ and our
variational parameter $\beta$ kept fixed~at,~say,~$\beta=1$:
\begin{equation}\psi_{0,00}(\bm{x})\propto\exp(-\mu\,r)\
,\qquad\widetilde\psi_{0,00}(\bm{p})\propto(\bm{p}^2+\mu^2)^{-2}\
.\label{Eq:GSTF}\end{equation}

\subsection{Accuracy and reliability of solutions: Master virial
theorem}\label{Subsec:ARS}Quality and accuracy
\cite{Lucha:Q&Aa,Lucha:Q&Ab} of an approximate solution to a
bound-state equation in use can be~easily scrutinized by a
relativistic \emph{generalization\/} \cite{Lucha:RVTa} of the
virial theorem: All eigenstates $|\chi\rangle$ of operators of the
form $T(\bm{p})+V(\bm{x})$ satisfy a master equation
\cite{Lucha:RVTb} relating the expectation values of radial
derivatives:\begin{equation}\left\langle\chi\left|\,\bm{p}\cdot
\frac{\partial\,T}{\partial\bm{p}}(\bm{p})\,\right|\chi\right\rangle
=\left\langle\chi\left|\,\bm{x}\cdot\frac{\partial\,V}{\partial\bm{x}}
(\bm{x})\,\right|\chi\right\rangle.\end{equation}

\subsection{Desperately seeking analytic results: Seductions and
pitfalls}\label{Subsec:DSAR}Aiming at \emph{analytic
approximations\/} to the exact solutions of spinless Salpeter
equations at (almost)~any price triggers hectic activity
\cite{Hpa,Hpb,Hpc,Hpd,Hpe,Hpf,Hpg}: Frequently, close encounters
with the nonlocality of the operator $H$ are avoided by expanding
$T(\bm{p})$ up to $O(\bm{p}^4/m^4),$ to deal with the apparently
nicer behaving operators\begin{equation}H_{\rm
p}\equiv2\,m+\frac{\bm{p}^2}{m}-\frac{\bm{p}^4}{4\,m^3}+V(\bm{x})\
.\end{equation}However, the expectation value of such
``pseudo-spinless-Salpeter Hamiltonian'' $H_{\rm p}$ over, for
example, the trial function (\ref{Eq:GSTF}), that is,
$\phi(r)\propto\exp(-\mu\,r),$ reveals that this operator $H_{\rm
p}$ is \emph{not\/} bounded from below:\begin{equation}
\left\langle H_{\rm p}\right\rangle=2\,m+\frac{\mu^2}{m}
-\frac{5\,\mu^4}{4\,m^3}+\left\langle V(\bm{x})\right\rangle\qquad
\Longrightarrow\qquad\lim_{\mu\to\infty}\left\langle H_{\rm
p}\right\rangle=-\infty\qquad\Longrightarrow\qquad E_0\le-\infty\
.\end{equation}Consequently, all searches for ground states must
be doomed to fail. However, a perturbative approach to
$\bm{p}^4/4\,m^3,$ adopted correctly, \emph{may\/} save one's day.
An expansion over potential-inspired functions \cite{Pekeris}
mitigates the singularity of the Laplacian's centrifugal term
$\propto r^{-2},$ but alters the full effective~potential.

\section{Application to potential regular at the origin:
Woods--Saxon problem \cite{Lucha:WS}}\label{Sec:AWSP}The
Woods--Saxon (WS) potential is a rather tame potential, familiar
from nuclear physics, determined by coupling strength $V_0,$
potential width $R,$ and surface thickness $a,$ all of them
assumed to be real~\cite{WSP}:\begin{equation}V(r)=V_{\rm
WS}(r)\equiv-\frac{V_0}{1+\exp\left(\frac{r-R}{a}\right)}\ ,\qquad
V_0>0\ ,\qquad R\ge0\ ,\qquad a>0\ .\end{equation}For
definiteness, let's impose the concepts in Sect.~\ref{Sec:ASSC},
as applicable, to the WS eigenvalue problem~\cite{Lucha:WS} for
the set of mass and potential parameter numerical values of Table
\ref{Tab:WSNPV}, dubbed ``physical'' in Ref.~\cite{Hamzavi_CPC}:

\begin{table}[b]\centering\caption{Numerical parameter values
adopted for the semirelativistic WS problem by the treatment of
Ref.~\cite{Hamzavi_CPC}.}\label{Tab:WSNPV}
\begin{tabular}{lcccc}\hline\hline\\[-1.5ex]
Parameter&$m$&$V_0$&$R$&$a$\\[1ex]\hline\\[-1.5ex]Numerical value
&$940.271\;\mbox{MeV}$&$67.70352\;\mbox{MeV}$&$7.6136\;\mbox{fm}$&
$0.65\;\mbox{fm}$\\[1ex]\hline\hline\end{tabular}\end{table}

\begin{itemize}\item The \emph{lower\/} limit to the energy
spectrum is, clearly, $E_0\ge\inf_rV(r)=V(0)=-67.70296\;\mbox{MeV}
\gtrapprox-V_0.$\item The energy interval defined by this bound,
$V(0)<E_k\le0$ ($k=0,1,2,\dots,N$), may accommodate for
relativistic and nonrelativistic kinematics, respectively,
$N\le850$ and $N\le1201$ eigenstates,~at~most.\item For (semi-)
relativistic WS bound states identified by radial and orbital
angular momentum quantum numbers $n_r$ and $\ell,$ Table
\ref{Tab:WSUBEL} presents variational \emph{upper\/} bounds to
their \emph{binding\/} energies derived for our setup
$\mu=1\;\mbox{GeV},$ $\beta=1,$ and subspace dimension $d=25,$ and
the bounds' Schr\"odinger~counterparts.\item The system
characterized by the parameter values in Table \ref{Tab:WSNPV}
hardly warrants its relativistic treatment since it is
\emph{highly\/} nonrelativistic, as the expectation value of
$\bm{p}^2/m^2$ over the first trial state (\ref{Eq:GSTF})~reveals:
\begin{equation}\left\langle\frac{\bm{p}^2}{m^2}\right\rangle
\approx6\times10^{-3}\ .\end{equation}\end{itemize}

\begin{table}[t]\centering\caption{Variational and nonrelativistic
\emph{upper\/} bounds to semirelativistic WS \emph{binding\/}
energies (in units of GeV).}\label{Tab:WSUBEL}
\begin{tabular}{cccc}\hline\hline\\[-1.5ex]$n_r$&$\ell$&Spinless
Salpeter equation&Schr\"odinger equation\\[1ex]\hline\\[-1.5ex]
0&0&$-0.06032$&$-0.06030$\\&1&$-0.05309$&$-0.05305$\\[.5ex]
1&0&$-0.04119$&$-0.04108$\\&1&$-0.02967$&$-0.02946$\\[.5ex]
2&0&$-0.01527$&$-0.01545$\\&1&$-0.00233$&$-0.00362$\\[1ex]
\hline\hline\end{tabular}\end{table}

\section{Application to a potential singular at the origin:
Hulth\'en problem}\label{Sec:AHP}The short-range Hulth\'en
potential is characterized by two parameters, coupling strength
$v$ and range~$b$:
\begin{equation}V(r)=V_{\rm H}(r)\equiv-\frac{v}{\exp(b\,r)-1}\
,\qquad b>0\ ,\qquad v\ge0\ .\label{Eq:VH}\end{equation}To
facilitate comparability, we phrase our remarks for the parameter
values used by Ref.~\cite{Zarrinkamar} (Table~\ref{Tab:HNPV}):
\begin{itemize}\item From the expectation value of the Hamiltonian
(\ref{Eq:H}) with Hulth\'en potential (\ref{Eq:VH}) over the trial
state~(\ref{Eq:GSTF}) we learn that its boundedness from below
requires the potential parameters to satisfy $v/b<16/(3\,\pi).$
\item Iff $\kappa\ge v/b,$ Hulth\'en's potential (\ref{Eq:VH}) is
bounded from below by any Coulomb potential $V_{\rm
C}(r)\equiv-\kappa/r.$ Thus, lower bounds to the relativistic
Coulomb problem, such as the one given by Herbst \cite{Herbst},
apply.\item For a \emph{Schr\"odinger\/} operator
(\ref{Eq:Schreq}) with Hulth\'en potential (\ref{Eq:VH}), the
eigenvalues for $\ell=0$ states read~\cite{Flugge}\begin{equation}
E_n=-\frac{\left(2\,\mu\,v-n^2\,b^2\right)^2}{8\,\mu\,n^2\,b^2}\
,\qquad n=1,2,3,\dots\ ,\qquad n^2\,b^2\le2\,\mu\,v\
.\label{Eq:NRHEV}\end{equation}\item Table \ref{Tab:HUBEL} lists,
for $\ell=0$ states, the \emph{upper\/} bounds to (semi-)
relativistic Hulth\'en \emph{binding\/} energies found
\emph{variationally\/} for $\mu=1,$ $\beta=1,$ and $d=25,$ or
represented by the \emph{analytically\/} given
eigenvalues~(\ref{Eq:NRHEV}).\end{itemize}

\begin{table}[b]\centering\caption{Numerical parameter values
adopted for the semirelativistic Hulth\'en problem in the study of
Ref.~\cite{Zarrinkamar}.}\label{Tab:HNPV}\begin{tabular}{lccc}
\hline\hline\\[-1.5ex]Parameter&$m$&$b$&$v$\\[1ex]\hline\\[-1.5ex]
Numerical value (arbitrary units)&$1$&$0.15$&$0.11$\\[1ex]
\hline\hline\end{tabular}\end{table}

\begin{table}[b]\centering\caption{Variational and Schr\"odinger
\emph{upper\/} bounds to semirelativistic Hulth\'en
\emph{binding\/} energies (arbitrary units).}\label{Tab:HUBEL}
\begin{tabular}{cccc}\hline\hline\\[-1.5ex]$n_r$&$\ell$&Spinless
Salpeter equation&Schr\"odinger equation\\[1ex]\hline\\[-1.5ex]
0&0&$-0.10577$&$-0.085069\dot4$\\[.5ex]
1&0&$-0.0022398$&$-0.00\dot1$\\[1ex]
\hline\hline\end{tabular}\end{table}

\section{Summary and conclusions}\label{Sec:S&C}Even though the
spinless Salpeter equation resists to being solved by analytical
techniques, a variety of elementary considerations allows us to
draw a pretty clear picture of the solutions to be expected~out of
such efforts. Nevertheless, \emph{not all\/} solutions offered in
the literature do respect the frame of this~picture.

\end{document}